\documentclass[aps,pra,reprint,superscriptaddress]{revtex4-2}

\usepackage{xcolor}

\usepackage{nicematrix}
\usepackage{graphicx}% Include figure files
\usepackage{dcolumn}% Align table columns on decimal point
\usepackage{bm}% bold math
\begin{document}

%\preprint{APS/123-QED}

\title{%Quantum Criticality in Dilute Hopfield Model\\
%Quantum Criticality in the Hopfield Model of Dilute Memories\\
%Characteristic Time Complexity in Quantum Hopfield Model\\
Quantum Hopfield Model with Dilute Memories}% Force line breaks with \\
%\thanks{A footnote to the article title}%

\author{Rongfeng Xie}
\affiliation{School of Physics and Astronomy, University of Minnesota, Minneapolis, Minnesota 55455, USA}
\author{Alex Kamenev}
\affiliation{School of Physics and Astronomy, University of Minnesota, Minneapolis, Minnesota 55455, USA}
\affiliation{William I. Fine Theoretical Physics Institute, University of Minnesota, Minneapolis, Minnesota 55455, USA}

\date{\today}% It is always \today, today,
             %  but any date may be explicitly specified

\begin{abstract}
We discuss adiabatic spectra and dynamics of the 
quantum, i.e. transverse field, Hopfield model with
dilute memories (the number of stored patterns $p < log_2 N$, where $N$ is the number of qubits). At some critical transverse field  the model undergoes the quantum phase transition from the 
ordered to the paramagnetic state. The corresponding critical exponents are calculated and used to determine the efficiency of quantum annealing protocols. We also discuss implications of these results for the quantum annealing  of generic spin glass models. 
\end{abstract}

%\keywords{Suggested keywords}%Use showkeys class option if keyword
                              %display desired
\maketitle

%\tableofcontents

\section{Introduction} \label{sec_1}
The Hopfield model \cite{hopfield_1982,farhat_1985,aiyer_1990} played a significant role in the early development of neural network theory, contributing to the understanding of the associative memory and the computational capabilities of interconnected neurons. Renowned for its applications in computational neuroscience, the Hopfield model takes its origin in the classical Ising model of statistical mechanics. Many of its key properties, e.g. the critical capacity $\alpha_c \approx 0.14$ \cite{amit_1985,daniel_1985}, have been evaluated using the replica trick and cavity method, borrowed from the theory of  disordered systems and spin glasses \cite{mezard_1987}. 

Recent rapid developments of the quantum computing and quantum information theory, have stimulated the introduction  of quantum mechanics into neural network models \cite{schuld_2014,mcclean_2018,beer_2020,sharma_2022,cao_2017}. One of the earliest attempts is the quantization of the Hopfield model by substituting classical binary bits with the spin-$\frac{1}{2}$ operators. This paper discusses the transverse field Hopfield model, defined as 
\begin{equation} \label{H_1_1}
    H = -\frac{1}{2}\sum_{i\neq j}^{N} J_{ij}\sigma_i^z \sigma_j^z - \Gamma \sum_{i=1}^{N} \sigma_i^x 
\end{equation}
where $\sigma_i^x$ and $\sigma_i^z$ are Pauli matrices and $\Gamma$ is a transverse field. The coupling matrix $J_{ij}$ encodes the information of $p$ binary patterns (or memories) by the Hebb's rule  \cite{hebb_2005}, 
\begin{equation}
    \label{Eq:Hopfield-J}
    J_{ij} = \frac{1}{N} \sum_{\mu = 1}^{p}\xi_i^{\mu} \xi_{j}^{\mu}
\end{equation}
where $ \{\xi_i^{\mu} \}$ with $\mu = 1\ldots ,p$, and $i=1,\ldots,N$ represent $p$ patterns, each $N$-bit long.  For a statistical analysis it is often convenient to take $ \{\xi_i^{\mu} \}$ as independent random variables, say  with the equal probability to be either $+1$ or $-1$.
 Free energy and the phase diagram of the model were studied in Refs.~\cite{nishimori_1996,shcherbina_2020}. Specifically, Nishimori \textit{et al}.~\cite{nishimori_1996} explicitly showed that the replica method can be applied in such a quantum model where the external parameter $\Gamma$ plays the role similar to temperature in the classical model (one could expect this considering the correspondence between the $d+1$ dimensional classical model and $d$ dimensional quantum model). The quantum phase diagram characterized by $\Gamma$ and $\alpha$ (=$p/N$) is  nearly the same as the classical one found in Ref.~\cite{amit_1985}, except that temperature is replaced with $\Gamma$. This similarity shows that quantum fluctuations, in influencing the macroscopic properties of the system, bear a striking resemblance to thermal fluctuations.

The dynamics of updating the neurons in the classical model, such as Glauber dynamics, have been explored in many aspects \cite{daniel_1985,bruce_1987,huang_2010}. The quantum model (\ref{H_1_1}), evolving according to the Schr\"odinger equation, may include time-dependent parameters, such as the transverse field $\Gamma(t)$.  Specific time-dependent protocols can be adopted from appropriate quantum algorithms, e.g., quantum annealing \cite{kadowaki_1998,brooke_1999,santoro_2002,santoro_2006,hauke_2020} or adiabatic quantum computing \cite{farhi_2000,Albash_2018}. These protocols,  often regarded as algorithms with potential quantum advantages in solving combinatorial optimization problems, have gained wide attention. Given that numerous combinatorial optimization problems can be recast as the energy minimization of an Ising-like Hamiltonian \cite{lucas_2014,wang_2022,zhang_2024} sharing the similar form of (\ref{H_1_1}), it is natural to investigate the quantum annealing problem associated with the Hopfield model. 
The model was also extended to the $k$-body interaction scenarios. In Ref.~\cite{seki_2015}, it was found that the antiferromagnetic transverse field allows one to choose an annealing path to avoid problematic first-order phase transitions for specific $k$'s. In open quantum systems \cite{inoue_2011,rotondo_2018}, the dynamics are described by Markovian processes, allowing for the simultaneous treatment of both thermal and quantum coherent effects. It was found that quantum fluctuations can lead to the emergence of unique non-equilibrium phases. While being mentioned in these extended models, the lack of discussion of quantum annealing in quantum Hopfield itself leads to this work. 

The quantum advantage of adiabatic quantum computation and quantum annealing was questioned early on, primarily attributing it to the first-order quantum phase transition in NP-hard problems in which exponentially small eigenvalue gaps appear in the adiabatic energy spectrum \cite{altshuler_2010,amin_2009}. 
In a recent study of the two-pattern Gaussian Hopfield model and the  Sherrington-Kirkpatrick (SK) model \cite{knysh_2016}, it was argued that there are $O(\log N)$ exponentially small gaps in the glassy phase.
Consequently, various strategies, such as quantum approximate optimization algorithms \cite{farhi_2014,farhi_2022}, reverse and iterative cyclic annealing \cite{ohkuwa_2018,wang_2022,zhang_2024}, counter-diabatic driving \cite{berry_2009,sels_2017,claeys_2019,hegade_2022} and quantum walks \cite{callison_2019}, have been devised in the non-adiabatic regime with the aim of mitigating these challenges. These protocols provide future directions for studying the quantum Hopfield model.

In the present work, we investigate the characteristic annealing time in the quantum Hopfield model. We develop our theory using a two-level Landau-Zener formula in the adiabatic regime. It turns out that two scaling exponents are crucial, i.e., the dynamical energy gap in the thermodynamic limit around the critical point ($\sim |\Gamma - \Gamma_c|^{a}$) and the finite-size scaling of the critical energy gap ($\sim N^{-b}$). A well-defined second-order phase transition point is confirmed in the case of $p=1$ by noticing that it shares the same form of the Lipkin-Meshkov-Glick (LMG) model \cite{glick_1965}. The critical behaviors of the LMG model were studied in \cite{Dusuel_2004,Leyvraz_2005}. Reference~\cite{Dusuel_2004} utilizes the Holstein-Primakoff boson representation along with the continuous unitary transformation technique to accurately compute finite-size scaling exponents for various properties such as the energy gap and ground state energy, while Ref.~\cite{Leyvraz_2005} introduces a purely semiclassical method to approach similar results. Reference~\cite{Caneva_2008} studied the adiabatic quantum dynamics of the LMG model, showing different scaling regimes for the residual energy. Following Ref.~\cite{Dusuel_2004}, we use the large-$N$ expansion of the Holstein-Primakoff transformation to evaluate the critical exponents $a$ and $b$.  This idea is further extended to the case of dilute memories ($p<log_2 N$), where we treat different clusters of $1/2$-spins as large spins. We explicitly show that these two exponents are not spoiled by dilute memories. They only form successive finite-size local minima near the critical point. One interesting fact is that $b = 1/3$ is postulated to be universal in infinite-ranged non-disorder quantum systems \cite{botet_1982}, where it has been shown that $b$ is closely related to the upper critical dimension of the quantum Ising model. For the regular Ising model of dimension $d$, the conventional notation for our $a$ and $b$ is $z\nu$ and $z/d$ respectively, where $\nu$ is the critical exponent of correlation length and $z$ is the dynamic critical exponent. One recovers our results ($a=1/2, b = 1/3$) by assuming $\nu=1/2$ (mean-field value), $z=1$ \cite{sachdev_2011} and $d=3$ (upper critical dimension for the quantum Ising model). The two-pattern Gaussian Hopfield model \cite{knysh_2016} and our study of the Hopfield model with dilute memories confirm $b=1/3$, suggesting that this result might be true for many spin glass systems. Thus, it further suggests our discussion of the characteristic annealing time ($O(N)$) is general, i.e. one may expect such a time scale in quantum annealing of other infinite-ranged spin glass models, e.g., in the diabatic regime where one can ignore exponentially small gaps.

The remainder of this paper is organized as follows. In Section \ref{sec_2}, we derive the characteristic annealing time for a generic quantum annealing problem, provided there is a second-order phase transition point with relevant scaling exponents. In Section \ref{sec_3} and Section \ref{sec_4}, we explicitly show the existence of such a second-order phase transition point in the quantum Hopfield model for $p < \log_2 N $. Finally, Section \ref{sec_5} summarizes our findings and provides an outlook for future work.

\section{The Quantum Annealing Problem} \label{sec_2} 

A quantum annealing problem is described by the Schrödinger equation
\begin{equation}
    i\partial_t | \psi(t) \rangle = H(t)|\psi(t)\rangle
\end{equation}
where the Hamiltonian changes along an annealing trajectory in its parameter space, ending at the Hamiltonian of the optimization problem at hand. For an adiabatic evolution the state of the system follows its instantaneous ground state, starting from a known one of the initial Hamiltonian, all the way to that of the problem of interest. 
The time scale, $T$, required by adiabaticity scales algebraically with the minimal energy gap, $\Delta$, between the ground and  a lowest excited state allowed by symmetries,  $T \sim  \Delta^{-\beta}$ \cite{Albash_2018}, where $\beta$ is a positive exponent.
Here we consider the simplest annealing protocol 
\begin{equation}
            \label{eq:ham1}
    H(t) = \Gamma(t) H_x + H_z ,
\end{equation}
where $\Gamma(t)$ is a monotonically decreasing function of time, and for the transverse field Hopfield model
\begin{equation}
\label{eq:ham2}
    H_x = -\sum_{i = 1}^{N}\sigma_i^x ; \quad \ \ \ H_z = - \frac{1}{2} \sum_{i\neq j}^N J_{ij}\sigma_i^z \sigma_j^z, 
\end{equation}
where $J_{ij}$ is the Hopfield coupling matrix (\ref{Eq:Hopfield-J}). 

If $\Gamma$ is considered as a static parameter, the ground state of the Hamiltonian (\ref{eq:ham1}), (\ref{eq:ham2}) in the thermodynamic limit, $N\to\infty$,  undergoes a continuous phase transition at  $\Gamma = \Gamma_c$, between the paramagnetic state at $\Gamma > \Gamma_c$ and the ordered state at $\Gamma < \Gamma_c$.
The former state is gaped, while the latter  exhibits a sequence of excited states, which become continuous in the $N\to \infty$ limit. The macroscopic gap, $g$, in the paramagnetic state scales as  
\begin{equation}
        \label{eq:g}
    g(\Gamma) \propto (\Gamma-\Gamma_c)^{a} \ ,
\end{equation}
with $a=1/2$. Our goal is to understand the finite $N$ fine structure of this picture, to estimate  how the annealing time should scale with $N$. 

\begin{figure}[!htbp]
\centering
    \includegraphics[width=\columnwidth]{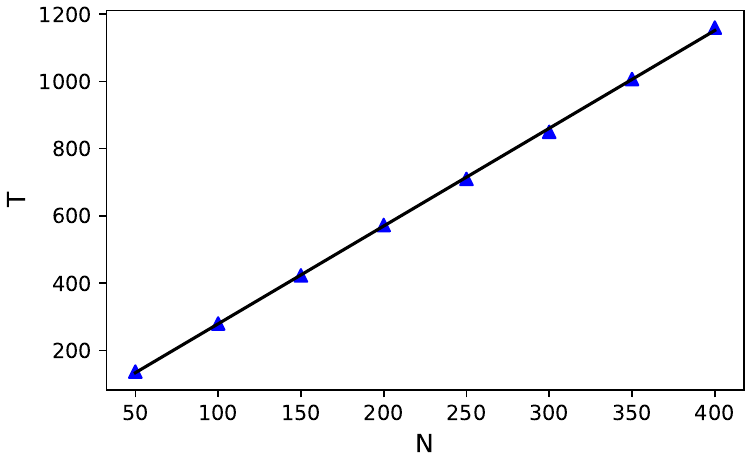}
    \caption{ \label{fig1} Annealing time needed to achieve 90\% success probability with respect to system size $N$(the case of $p=1$). The success probability is defined as the probability of finding the ground state after annealing.  Due to the symmetry explained in section \ref{sec_3}, the reduced Hilbert space is of dimension $O(N)$, which allows one to compute the evolution operator for hundreds of qubits.}
\end{figure}

We will show that in the ordered phase $\Gamma<\Gamma_c$, there is a sequence of $p$ avoided crossing transitions between the ground and low energy states, describing Hopfield patterns. The characteristic distance between these avoided crossings scales as $\delta \Gamma \sim (pN)^{-1/2}$, while the finite size energy gaps, $\Delta$, at the transitions scale  as    
\begin{equation}
\Delta \propto N^{-b} \ ,
\end{equation}
with $b=1/3$, which is much less than the separation between the transitions, $\delta \Gamma$, in the large $N$ limit. As a result, the individual transitions maintain their 
respective identities and the corresponding probability, $P$, of the {\em non-adiabatic} dynamic transitions can be estimated using 
the Landau-Zener formula \cite{Vitanov_1999}
\begin{equation}
    P = e^{-2\pi D} \ \ , \ \ D = \frac{\Delta^2}{4|\partial g(\Gamma)/\partial t|} .
\end{equation}
Although one can optimize the $\Gamma(t)$ protocol 
(often referred to as the quantum adiabatic brachistochrone \cite{Rezakhani_2009}), we will assume 
the simplest linear schedule, $\Gamma\sim t$, for the most conservative estimate. In this case  the characteristic annealing time $T$ scales as 
\begin{equation}
    T\sim \frac{|\partial g(\Gamma) / \partial \Gamma|}{\Delta^2} \ .
\end{equation}
Since the diabatic transitions only happen when $g \sim \Delta$, i.e. in the parameter window $|\Gamma-\Gamma_c| \sim \Delta^{1/a} $,  the relevant slope $|\partial g(\Gamma) / \partial \Gamma| \propto |\Gamma-\Gamma_c|^{a - 1}$ is estimated as $\Delta^{1-\frac{1}{a}}$. Consequently, the annealing time scales as 
\begin{equation}
    T \sim \Delta^{-1 - \frac{1}{a}}, 
\end{equation}
or in terms of the system size, $N$, as 
\begin{equation}
    T \sim N^{b(1+\frac{1}{a})}.
\end{equation}
Substituting $b = 1/3$ and $a = 1/2$, one estimates the characteristic annealing time as $T \sim N$. Numerical calculations are made for the single pattern case, and the linearity between $T$ and $N$ is clearly seen in Fig.~\ref{fig1}.

It is worth mentioning that the global $Z_2$ symmetry exists in such a two-body interaction model, associated with the parity operator
\begin{equation}
    Z  = \sigma_1^x \otimes \sigma_2^x \otimes ...\otimes \sigma_N^x \equiv \prod_{i} \sigma_i^x
\end{equation}
 whose effect is equivalent to flipping all spins on computational bases. In particular,  the initial state (the ground state of $H_x$) is even
 \begin{equation}
     Z|+++...+++\rangle = |+++...+++\rangle
 \end{equation}
 where $\sigma_x|+\rangle = |+\rangle$. Thus, the dynamical transition only happens among states with even parity and the energy gap $\Delta$ of interest refers to the gap between the ground state and the second excited state (i.e the first excited dynamical state with the same parity as the ground state).

\section{$p=1$ case} 
                \label{sec_3}
                
As a warm up exercise let us consider the case of a single energy minimum, $p=1$.  Without loss of generality we set this minimum to be a ferromagnetic state, encoded by $\xi^{(1)} = \{1,1,1,...,1\}$.  The Hamiltonian acquire a simple form 
\begin{equation} 
H = - \frac{1}{2N} \left(\sum_{i = 1}^{N} \sigma_i^z\right)^{\!2} - \Gamma \sum_{i=1}^{N}\sigma_i^x
\end{equation}
where the constant term is omitted.  This is the LMG model \cite{glick_1965}, defined through spin-$N$ operators:
\begin{equation}
S_x = \frac{1}{2}\sum_{i=1}^N \sigma_i^x \ \ , \ \ S_z = \frac{1}{2}\sum_{i=1}^N \sigma_i^z
\end{equation}  
which satisfy SU(2) algebra. The Hamiltonian acquires the form 
\begin{equation} \label{H-LMG}
H = -\frac{2}{N}\, S_z^2 - 2\Gamma S_x.
\end{equation}
This shows that the $2^N$-dimensional Hilbert space splits onto blocks labeled by the total spin $S=0,1,\ldots N/2$ (hereafter we assume even $N$ to simplify the notations). In particular, the ferromagnetic ground state belongs to the maximal spin $S=N/2$ block, which has multiplicity one. Therefore the relevant subspace is $2S+1=N+1$-dimensional, which allows us to simulate time-dependent Schr\"odinger evolution for very large $N$.          

\begin{figure}[!htbp]
\centering
    \includegraphics[width=\columnwidth]{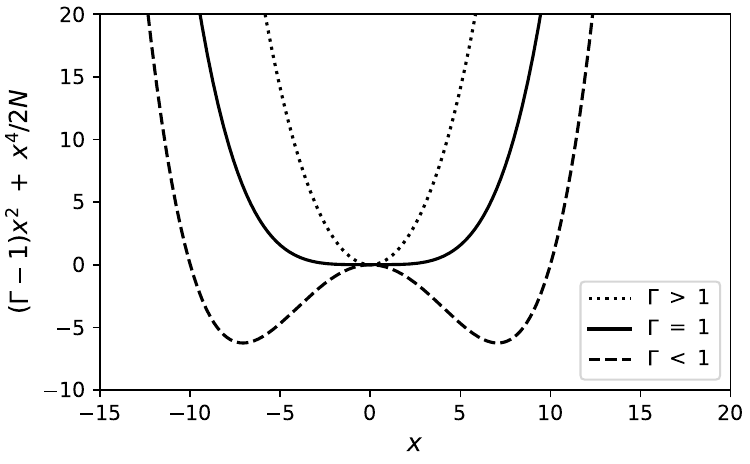}
    \caption{ \label{fig2} Transition from the harmonic potential to the double-well potential ($N=100$). }
\end{figure}

 The critical behavior of the LMG model was understood in Refs.~\cite{Dusuel_2004,Leyvraz_2005}. Here we adopt and simplify the treatment of Ref.~\cite{Dusuel_2004}, which will allow us to extend it to the $p>1$ Hopfield model.  To this end we employ the Holstein-Primakoff transformation \cite{auerbach_2012} for $SU(2)$ spin operators in the following form 
\begin{subequations}
\label{HPT}
\begin{equation}
S^x = S - a^{\dagger}a = \frac{N}{2} - a^{\dagger}a
\end{equation}
\begin{equation}
S^+ = \sqrt{2S-a^\dagger a}\ a = N^{\frac{1}{2}} \left( 1-\frac{a^\dagger a}{N} \right)^{\! \frac{1}{2}}a
\end{equation}
\begin{equation} 
S^- = a^\dagger \sqrt{2S-a^\dagger a} = N^{\frac{1}{2}}a^\dagger \left(1-\frac{a^\dagger a}{N} \right)^{\! \frac{1}{2}},
\end{equation}
\end{subequations}
where $S^\pm = S^z\mp i S^y$, and $a,a^\dagger$ are bosonic creation, annihilation operators with the commutation relation $[a,a^\dagger]=1$. Expanding the square roots here to the order of ${1}/{N}$, one finds (up to an additive constant) 
\begin{equation} \label{s3_H}
H = (2\Gamma-1)a^\dagger a - \frac{a^{\dagger2}+a^2}{2} + \frac{(a^\dagger+a)a^\dagger (a^\dagger+a)a +h.c.}{4N}.
\end{equation}
Reference \cite{Dusuel_2004} then used the continuous unitary transformation technique to proceed. In contrast, we find it more convenient to write this Hamiltonian via the canonical position and momentum operators, defined as  
\begin{equation}
        \label{phase-space}
a = \frac{1}{\sqrt{2}} (x+ip); \ \ \  \ \ \ a^\dagger = \frac{1}{\sqrt{2}} (x-ip).
\end{equation}
With their help the Hamiltonian (\ref{s3_H}) takes the form:
\begin{equation} \label{phi4}
H \approx \Gamma\, p^2 + (\Gamma -1)\, x^2 + \frac{x^4}{2N},
\end{equation}
where the involved approximations will be discussed shortly. 
For $\Gamma>\Gamma_c=1$ (i.e. in the paramagnetic phase) the quadratic part of this Hamiltonian 
represents a harmonic oscillator with the frequency
\begin{equation} \label{s3_H0}
\omega(\Gamma) = 2 \sqrt{\Gamma (\Gamma - 1)}.
\end{equation}
It provides a gap (\ref{eq:g}) between the ground state and the lowest excited state within the $S=N/2$ subspace (only states within this subspace can undergo dynamical Landau-Zener transitions) in the paramagnetic phase, and illustrates the corresponding exponent, $a = 1/2$.  
For $\Gamma<1$  Eq.~(\ref{phi4}) represents the double-well potential, (see Fig.~\ref{fig2}), with two closely spaced low-energy states. These two states are given by symmetric and antisymmetric superpositions (i.e. cat states) of all spin-up and all spin-down ferromagnetic states. In the ferromagnetic state, these two states are exponentially close in a parameter $\propto N(1-\Gamma)^{3/2}\Gamma^{-1/2}$, showing that they are exactly degenerate in the thermodynamic limit, $N\to\infty$.

\begin{figure}[!htbp]
\centering
    \includegraphics[width=\columnwidth]{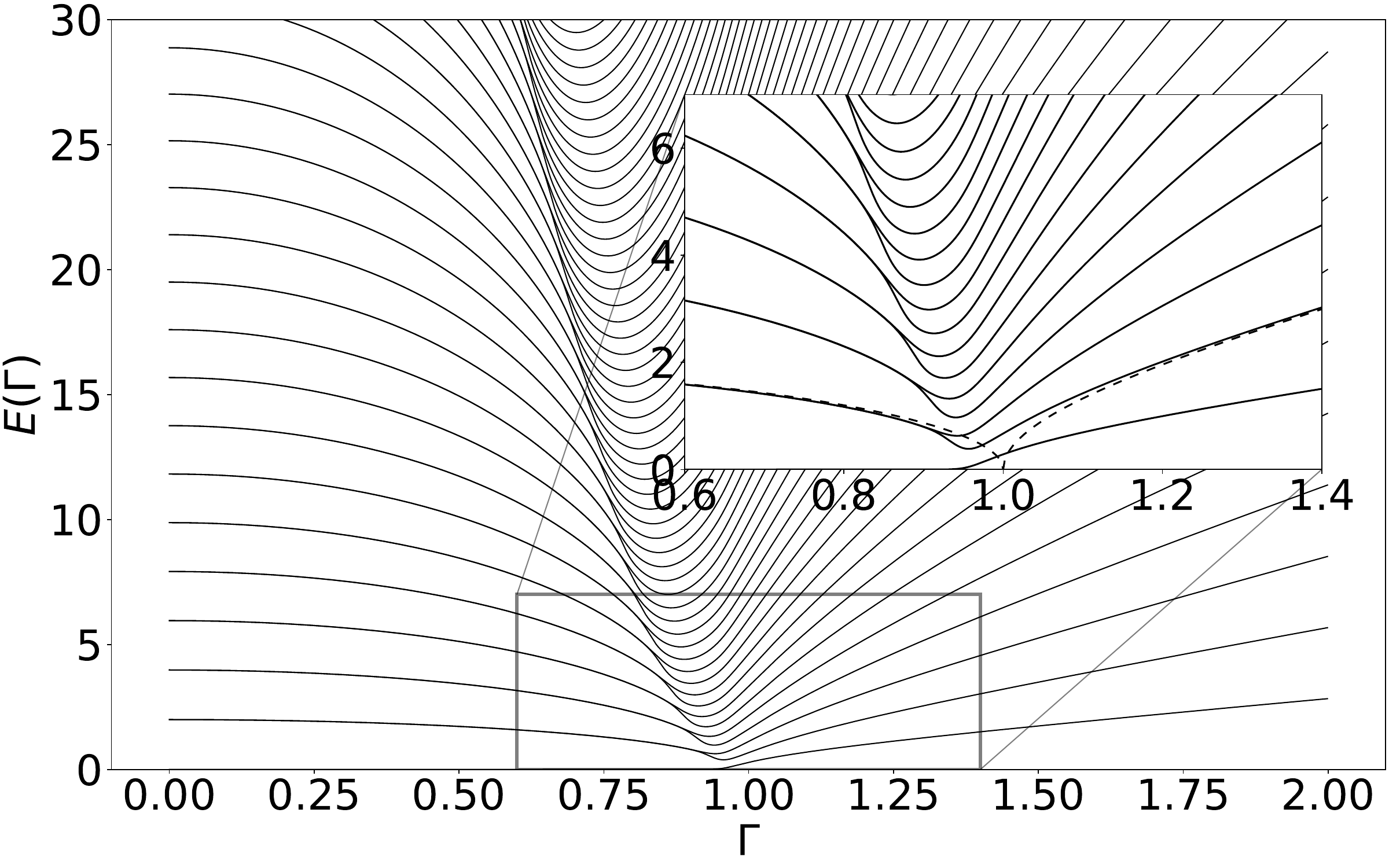}
    \caption{ \label{fig3}  $ S=N/2$ energy levels, $E_k-E_0$,  as functions of $\Gamma$ for $N=400$. The inset zooms onto the low energy part. Notice that energy levels become pair-wise degenerate (up to an exponentially small splitting) upon entering the ferromagnetic phase.   The second level in the thermodynamic limit (dashed in the inset), follows the $E_2(\Gamma)-E_0(\Gamma)\propto |\Gamma-\Gamma_c|^{1/2}$ behavior near $\Gamma_c$. The finite size energy gap scales as $E_2(\Gamma_c)-E_0(\Gamma_c)\propto N^{-1/3}$, as confirmed in  Fig.~\ref{fig4}.}
\end{figure}

\begin{figure}[!htbp]
\centering
    \includegraphics[width=\columnwidth]{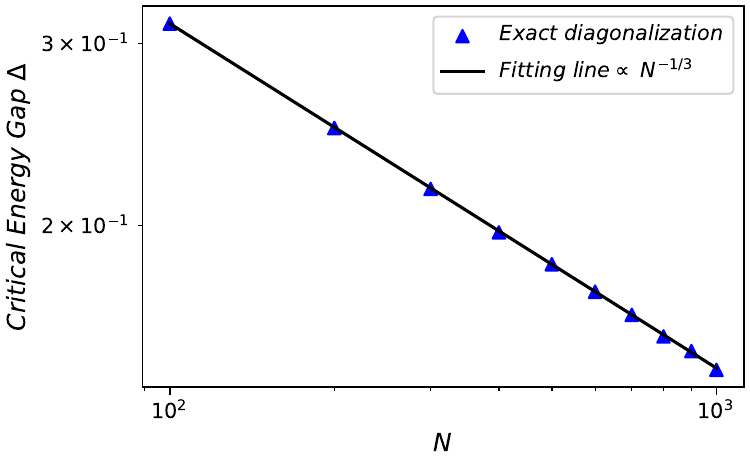}
    \caption{ \label{fig4} The log-log plot of the finite size gap $\Delta$ at $\Gamma=\Gamma_c$ vs the number of qubits ranging from 100 to 1000. }
\end{figure}

This shows that $\Gamma_c=1$ is the critical point, where  the energy gap closes in the thermodynamic limit. We are interested in the finite size scaling of such a critical gap. To investigate it, we notice that close to the criticality, where $0<\Gamma-1\ll 1$, for the oscillator energy $\sim \omega(\Gamma)$ the characteristic scales of the coordinate and momentum are $x\sim (\Gamma-1)^{-1/4}$ and $p\sim (\Gamma-1)^{1/4}$. Therefore, close to the criticality $x\gg p$ and therefore $a\approx a^\dagger\approx x/\sqrt{2}$. This is why going from Eq~(\ref{s3_H}) to Eq.~(\ref{phi4}) we kept only $x$ but not $p$ in the non-linear, $\sim 1/N$, terms. Focusing then at $\Gamma=\Gamma_c=1$, one arrives at the effective unharmonic oscillator  
\begin{equation}
H = p^2 + \frac{x^4}{2N}.
\end{equation}
Applying the Bohr-Sommerfeld quantization condition 
\begin{equation}
    \oint pdx = \int\! \sqrt{E_k - \frac{x^4}{2N}}\,\, dx = 2\pi (k+1/2),
\end{equation}
one notices that $E_k^{1/2} (E_kN)^{1/4}\propto k$ and thus 
\begin{equation}
    E_k - E_0 \propto N^{-{1}/{3}} k^{{4}/{3}} ,
\end{equation}
where integer $k$ labels the energy levels with energy $E_k$. The spectrum within the $S=N/2$ subspace is shown in Fig.~\ref{fig3}, and the exponent $b=1/3$ is confirmed in the numerical calculation, see Fig.~\ref{fig4}. Therefore, the finite-size energy gap  $\Delta = E_2 - E_0$ for the first excited dynamical state at the critical point is 
\begin{equation}
    \label{Delta}
    \Delta \propto N^{-1/3}.
\end{equation}

\section{$p<\log_2(N)$ case} \label{sec_4}

%\subsection{Rearrange the system into large spins}
For the single minimum, $p=1$, we have seen that the full Hilbert space with the dimensionality $2^N$ is factorized down to the block with dimensionality $N+1$. The reason behind this is that the Hamiltonian is invariant under a permutation of the spins, such that it can be written in the form of a LMG model. As a result, the total spin $S$ is a conserved quantity. 
For $1< p <\log_2N$ some of the permutation symmetries are preserved, i.e., the system can be regarded as a collection of large spins. For example, $p=2$: without loss of generality, one takes the first memory to be  $ \xi^{(1)} = \{1,1,...,1,1\} $, while the second one  
$\xi^{(2)} = \{ 1,..,1,-1,..,-1 \}$ with $N_1$\ positive units and $N_2$\ negative units ($N = N_1+N_2$). Though $\xi^{(2)}$ is random in general, such a choice is always allowed due to the permutation symmetry. Then, the Hamiltonian $H_z$ can be reduced to $-\frac{2}{N} [ (S_1^z+S_2^z)^2 + (S_1^z-S_2^z)^2]$, where 
$S_l^z $ is the $z$ component of the $l$'th large spin with the total spin $N_l/2$. Following the same procedure, for $p = 3$, one finds  
$H_z = -\frac{2}{N} [ (S_1^z+ S_2^z + S_3^z+S_4^z)^2 + (S_1^z+ S_2^z - S_3^z-S_4^z)^2 +(S_1^z -S_2^z + S_3^z-S_4^z)^2]$. Figure~\ref{fig5} explains  this procedure.
The resulting Hamiltonian acquires the form  $H = H_z +\Gamma H_x$, with 
\begin{subequations} 
\begin{equation} \label{H_4_1}
    H_z =- \frac{2}{N}\sum_{\mu = 1}^{p}\sum_{l,m=1}^{2^{p-1}}K_l^{\mu} K_m^{\mu}S_l^z S_m^z 
\end{equation}
\begin{equation}
    H_x = -2 \sum_{l=1}^{2^{p-1}} S_l^x ,
\end{equation}
\end{subequations}
where the large spins are defined as
\begin{equation}
    S_l^z = \frac{1}{2} \sum_{l_i}^{N_l} \sigma_{l_i}^{z} \ \ \ , \ \ \ S_l^x = \frac{1}{2} \sum_{l_i}^{N_l} \sigma_{l_i}^{x} .
\end{equation}
Here, the $p$ vectors $K^{\mu}$'s are the fixed strings of the length $2^{p-1}$, e.g., for $p = 3$ they are: $K^1 = \{1,1,1,1 \}$, $K^2 = \{1,1,-1,-1 \}$, and $K^3 = \{1,-1,1,-1\}$. 
\begin{figure}[h]
\centering
\begin{NiceTabular}{ccc}[hvlines]
 $p=1$  &  $ \xi^{(1)} =  \underbrace{ \uparrow \uparrow\uparrow ...\uparrow \uparrow\uparrow}_{N} $ & $K = \{1 \}$ \\
\Block{2-1}{$p=2$} & $\xi^{(1)} =\underbrace{ \uparrow ...\uparrow}_{N_1} \underbrace{ \uparrow ...\uparrow}_{N_2}   $   & $K^1 = \{1,1\}$   \\
                    & $\xi^{(2)} =\underbrace{ \uparrow ...\uparrow}_{N_1} \underbrace{ \downarrow ...\downarrow}_{N_2}   $   & $K^2 = \{1,-1\}$\\
\Block{3-1}{$p=3$} & $\xi^{(1)} = \underbrace{ \uparrow ...\uparrow}_{N_1} \underbrace{ \uparrow ...\uparrow}_{N_2} \underbrace{ \uparrow ...\uparrow}_{N_3} \underbrace{ \uparrow ...\uparrow}_{N_4}$ &  $K^1 = \{ 1,1,1,1\}$ \\
                    & $\xi^{(2)} = \underbrace{ \uparrow ...\uparrow}_{N_1} \underbrace{ \uparrow ...\uparrow}_{N_2} \underbrace{ \downarrow ...\downarrow}_{N_3} \underbrace{ \downarrow ...\downarrow}_{N_4}$ & $K^2 = \{ 1,1,-1,-1\}$ \\
                    & $\xi^{(3)} = \underbrace{ \uparrow ...\uparrow}_{N_1} \underbrace{ \downarrow ...\downarrow}_{N_2} \underbrace{ \uparrow ...\uparrow}_{N_3} \underbrace{ \downarrow ...\downarrow}_{N_4}$&  $K^3 = \{ 1,-1,1,-1\}$ 
%\Block{1-3}{$\vdots$}
\end{NiceTabular}
\caption{\label{fig5} Definition of $p$ mutually orthogonal vectors $K^\mu$ for an arbitrary $p$. For a given $p$, the $N_l$'s are integers with the constraint $\sum_{l=1}^{2^{p-1}}N_l = N$. }
\end{figure}

Performing the Holstein-Primakoff transformation, given by Eq.~(\ref{HPT}),  for each of $2^{p-1}$ spins of size $N_l/2$
and keeping the first order terms in $1/N_l$, one finds
\begin{equation}
H_x = 2\Gamma \sum_{l=1}^{2^{p-1}}a_l^\dagger a_l - \Gamma N
\end{equation}
\begin{eqnarray}
H_z = -\frac{1}{2N} \sum_{\mu = 1}^{p}\sum_{l,m=1}^{2^{p-1}}K_l^\mu K_m^\mu \sqrt{N_l N_m}  \nonumber \\
\times \left(a_l^\dagger +a_l - \frac{a_l^\dagger(a_l^\dagger + a_l)a_l }{2N_l}\right) \nonumber \\
\times \left(a_m^\dagger +a_m - \frac{a_m^\dagger(a_m^\dagger + a_m)a_m }{2N_m} \right) .
\end{eqnarray}
Separating the quadratic and quartic terms, denoted as $H_0$ and $H_1$ respectively, one finds 
\begin{equation} \label{s4_H0}
H_0 = 2\Gamma \sum_{l=1}^{2^{p-1}}a_l^\dagger a_l - \frac{1}{2} \sum_{\mu = 1}^{p}\sum_{l,m=1}^{2^{p-1}}  \tilde{K}_l^\mu \tilde{K}_m^\mu 
(a_l^\dagger +a_l)(a_m^\dagger +a_m);
\end{equation}
\begin{eqnarray} \label{s4_H1}
&&\!\!\!\!\!\! H_1 = \frac{1}{4N} \sum_{\mu = 1}^{p}\sum_{l,m=1}^{2^{p-1}} K_l^\mu K_m^\mu \left[\sqrt{\frac{N_l}{N_m}}\ (a_l^\dagger +a_l) a_m^\dagger \right. \nonumber \\ 
&&\!\!\!\!\!\!\! \left.
\times (a_m^\dagger +a_m) a_m  
 + \sqrt{\frac{N_m}{N_l}}\ a_l^\dagger(a_l^\dagger +a_l) a_l (a_m^\dagger +a_m) \right]\!,
\end{eqnarray}
where $\tilde{K}_l^\mu = \sqrt{\frac{N_l}{N}}K_l^\mu$.

%\subsection{Diagonalization of the quadratic Hamiltonian}
To proceed, let us define matrices,
\begin{equation}
    \label{KMT}
%(\textbf{K})_{\mu l} = K_l^{\mu}, 
( \tilde{ \textbf{K} }) _{\mu l} =  \tilde{K} _l^{\mu}; \qquad 
\bf M = \tilde{K}^T \tilde{K}; \qquad T = \tilde{K} \tilde{K}^T,
\end{equation}
with the dimensions $p\times 2^{p-1}$, $\,2^{p-1}\times 2^{p-1}$ and $p\times p$ respectively. 
Since $\bf M$ is a real symmetric matrix, it may be diagonalized 
with the orthogonal rotation 
${\bf O^T M O}  = \mathrm{diag}\{ \epsilon_k \} $, where $k=1,\ldots, 2^{p-1}$. Only $p$ out of $2^{p-1}$ eigenvalues $\epsilon_k$ are non-zero, since the matrix $\bf M$ has a rank $p$. These $p$ non-zero $\epsilon_k$'s are eigenvalues of the $p\times p$ symmetric matrix $\bf T$. 
Upon orthogonal rotation of the operators: 
\begin{equation}\label{s4_U}
a_l = \sum_{m=1}^{2^{p-1}} {\bf O}_{lm} b_m;\quad  \ \ a_l ^\dagger  = \sum_{m=1}^{2^{p-1}} b_m^\dagger ({\bf O^{\mathrm{T}}})_{ml},
\end{equation}
the quadratic part of the Hamiltonian takes the form 
 \begin{equation}
H_0 = \sum_{k=1}^{2^{p-1}}\left[2\Gamma\, b_k^\dagger b_k -\frac{1}{2}\, \epsilon_k (b_k^\dagger + b_k)^2 \right].
 \end{equation}
As before, it describes a collection of harmonic oscillators  with frequences  
\begin{equation} \label{s4_omega}
\omega_k(\Gamma) = 2\sqrt{\Gamma (\Gamma - \epsilon_k)},
\end{equation}
for $\Gamma>\epsilon_k$.
The criticality is characterized by $p$ nearly degenerate low-energy modes, originating from $p$ eigenvalues of the $p\times p$ matrix $\bf T$.   

\begin{figure}[!htbp]
\centering
    \includegraphics[width=\columnwidth]{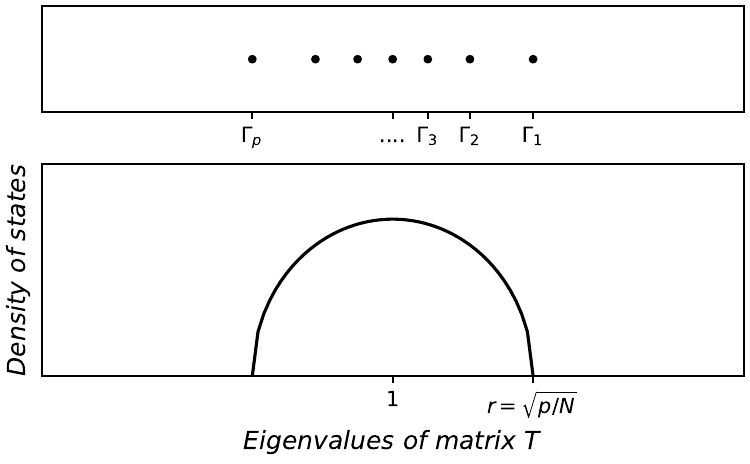}
    \caption{ \label{fig6} The upper panel visualizes the distribution of the eigenvalues of $\textbf{T}$  ($\Gamma_k$'s are non-zero $\epsilon_k$'s). The lower panel shows the density.  }
\end{figure}

To better understand the $\textbf{T}$ matrix, given by Eq.~(\ref{KMT}), consider the $p=3$ case. By virtue of the orthogonality of the $K^\mu$ vectors, the diagonal elements are all equal to unity, while the off-diagonal ones are composed of $N_l$'s with an equal number of positive and negative signs: 
\begin{equation} \label{s4_T}
    \textbf{T} = \begin{pmatrix}
        1 & \frac{N_1+N_2-N_3-N_4}{N} & \frac{N_1-N_2+N_3-N_4}{N} \\
        \frac{N_1+N_2-N_3-N_4}{N} & 1 & \frac{N_1-N_2-N_3+N_4}{N} \\
        \frac{N_1-N_2+N_3-N_4}{N} & \frac{N_1-N_2-N_3+N_4}{N} & 1
    \end{pmatrix}.
\end{equation}
Assuming random Hopfield minima $\xi^\mu$, one may argue that the off-diagonal elements are random numbers with  
\begin{equation}
\mathrm{mean}(T_{\mu \neq \nu}) = 0; \quad  \ \ \sigma^2(T_{\mu \neq \nu}) = \frac{1}{4N}. 
\end{equation}
Furthermore, in the limit $N\to \infty$ these off-diagonal elements are statistically independent, up to the symmetry. As a result, one can think of the ${\bf T}$-matrix as the unit matrix plus a random Gaussian orthogonal matrix. Its spectral density is therefore given by a Wigner semicircle \cite{mehta_2004} centered around one with the radius    
\begin{equation} \label{r_4_3}
r = 2\sqrt{p\sigma^2} = \sqrt{{p}/{N}}\ll 1 \  ,
\end{equation}
see Fig.~\ref{fig6}. The typical distance $d$ between the eigenvalues is thus 
\begin{equation} \label{d_4_3}
d \approx   \frac{2r}{p} =2 \sqrt{\frac{1}{Np}}  \sim \sqrt{\frac{1}{N}} \ .
\end{equation}
In the thermodynamic limit, the radius vanishes and $p$ degenerate zero modes appear at the critical point.

To understand the finite size behavior of the model we focus on these $p$ oscillators with $\epsilon_k\neq 0$.  Passing to the phase-space representation, as in Eq.~(\ref{phase-space}), and 
taking into account that near criticality $x_k\gg p_k$, 
one obtains (see Appendix A for details) the following Hamiltonian for the $p$ critical degrees of freedom: 
\begin{equation} \label{H'_4_4}
H =  \sum_{k=1}^{p}\left[\Gamma \,{p_k^2} + (\Gamma-\Gamma_k)\,{x_k^2} + \frac{x_k^4}{2N}\right] + \frac{3}{2N} \sum_{k\neq l}^{p}x_k^2 x_l^2,
\end{equation}
where $\Gamma_k$'s are defined as the non-zero $\epsilon_k$'s and are arranged such that $ \Gamma_p<...\Gamma_2 < \Gamma_1$. The first term of (\ref{H'_4_4}) represents $p$ independent non-linear oscillators, discussed in section \ref{sec_3}, and the second term represents their interactions. Without such interactions, the ground state is the state in which all oscillators are  in their ground states, while the low lying excited states are the states where one mode is excited and all other modes remain in their ground state. As  $\Gamma$ is decreased, the avoided crossings are  encountered at $\Gamma=\Gamma_1$, $\Gamma_2$,...,$\Gamma_p$.  
The corresponding finite-size gaps are given by Eq.~(\ref{Delta}). 

\begin{figure}[!htbp]
\centering
    \includegraphics[width=\columnwidth]{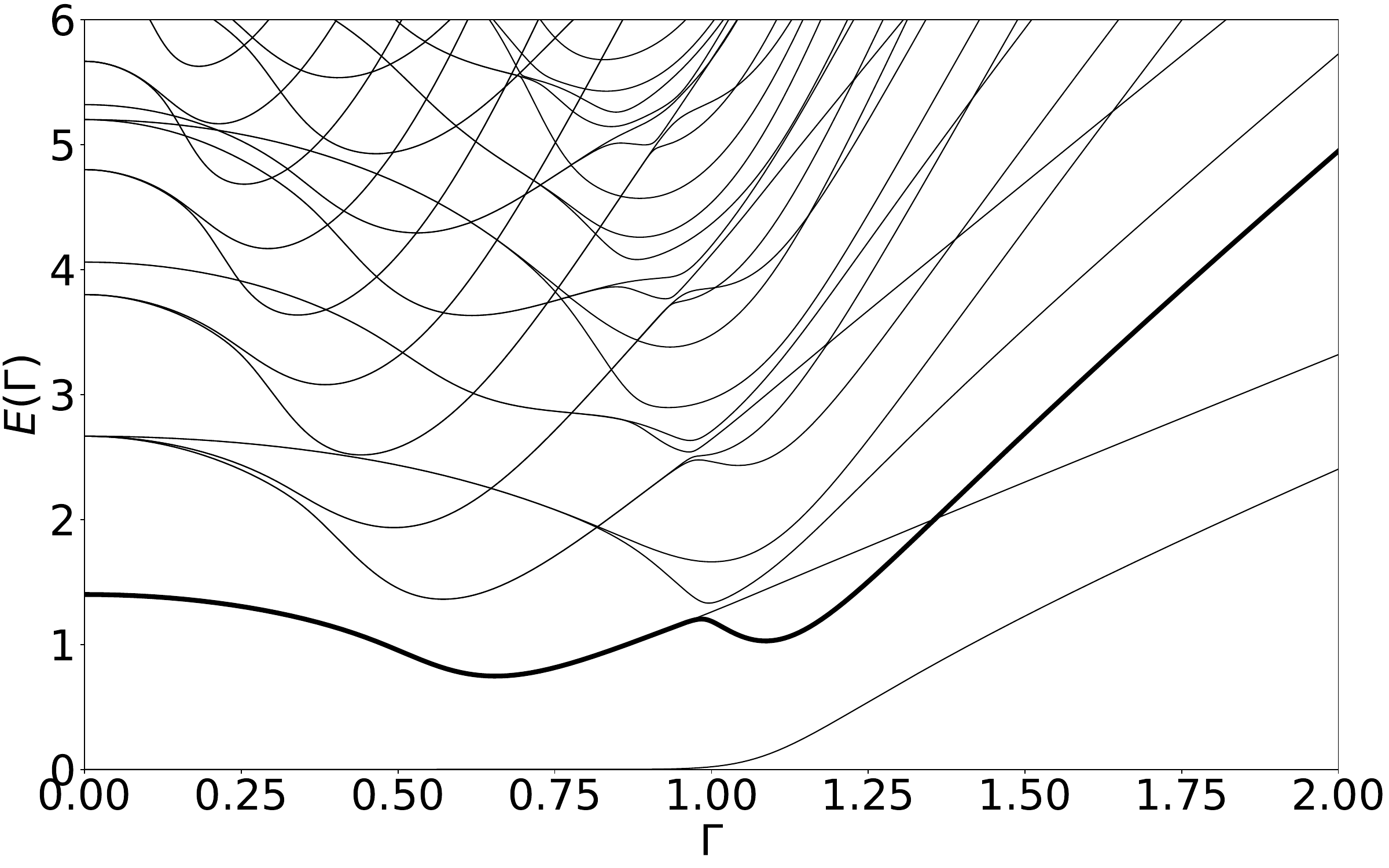}
    \caption{ \label{fig7} The energy levels for the case $p=2$ ($N=60$ with $N_1=40$, $N_2 =20$). The thickened black line is the first even state that is dynamical. Note that the original Hopfield model for $p=2$ has an additional symmetry ($Z_2\times Z_2$) that Eq.~(\ref{H_4_1}) $H_z = -\frac{4}{N}\left[(S_1^z)^2 + (S_2^z)^2 \right]$ has no interaction term. To present the general case, the interacting term $\sim S_1^z S_2^z$ was added manually to remove this symmetry in this numerical calculation.}
\end{figure}

An important question is if the intermode interactions in 
Eq.~(\ref{H'_4_4}) modify the critical exponents. For example, the exponents may acquire corrections in the small parameter $p/\log_2 N<1$. Here we argue that this is not the case. 
Indeed, for low energy excitation, the interaction term can be understood in the spirit of mean-field approximation
\begin{equation}
\frac{3}{2N} \sum_{k,l;k\neq l}^{p} x_k^2 x_l^2  \approx   \sum_{k}^{p} \delta \Gamma_k \,{x_k^2} \ ,
\end{equation}
\begin{equation} \label{s4_deltagamma}
\delta \Gamma_k = \frac{3}{N} \sum_{l \neq k}^{p} \langle  x_l^2 \rangle \ ,
\end{equation}
where the expectation $\langle \dots \rangle$ is taken with respect to the ground state without interactions. This bring the non-interacting Hamiltonian
\begin{equation} \label{H''_4_4}
H = \sum_{k=1}^{p} \left[ \Gamma\, {p_k^2} + (\Gamma - \Gamma_k^{'})\, {x_k^2} +\frac{x_k^4}{2N}  \right]  
\end{equation}
where $\Gamma_k^{'}= \Gamma_k - \delta \Gamma_k$. We found that the largest correction $\delta \Gamma_{k,max} = \delta \Gamma_p \sim \frac{1}{\sqrt{N}} $, which suggests that the average distance between the $\Gamma_k^{'}$'s is not qualitatively different from bare $\Gamma_k$'s. A careful calculation of $\delta \Gamma_k$ is detailed in Appendix \ref{ap_b}. In Appendix \ref{ap_c}, we further show that the typical finite size distance between the successive avoided crossings, given by Eq.~(\ref{d_4_3}), is large enough that all local gaps are well separated to maintain their identity along with the scaling, given by Eq.~(\ref{Delta}); see Fig.~\ref{fig7} for an example of $p=2$. This shows that rather than changing the critical exponents $b = 1/3$ and $a = 1/2$, the intermode interactions lead to effects which are sub-leading in higher powers of $1/N$. 

We thus conclude that for $p<p_c$ the finite size scaling of the Landau-Zener gaps remains to be given by Eq.~(\ref{Delta}), i.e. $b = \frac{1}{3}$. A value of the critical Hopfield capacity $p_c$ can not be determined from our considerations. It may scale logarithmically with the system size, $p_c\sim O(\log_2 N)$, or algebraically $p_c\sim O(N^\sigma)$. For $p<p_c$, the only effect of an increasing number of Hopfield minima, $p$, is an increase of the number of successive avoided crossings, encountered along the annealing path up to $p$. On the other hand, the scaling of individual gaps with $N$ continues to follow that of the $p=1$ LMG model.

\begin{figure*}
\centering
    \includegraphics[width=.8\columnwidth]{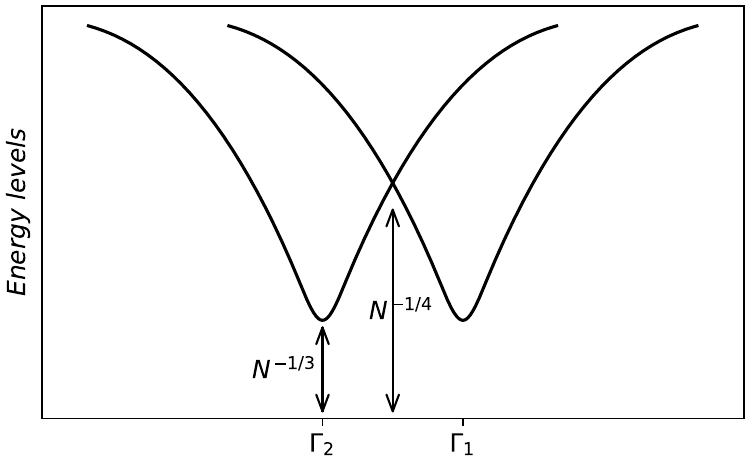}
    \includegraphics[width=.8\columnwidth]{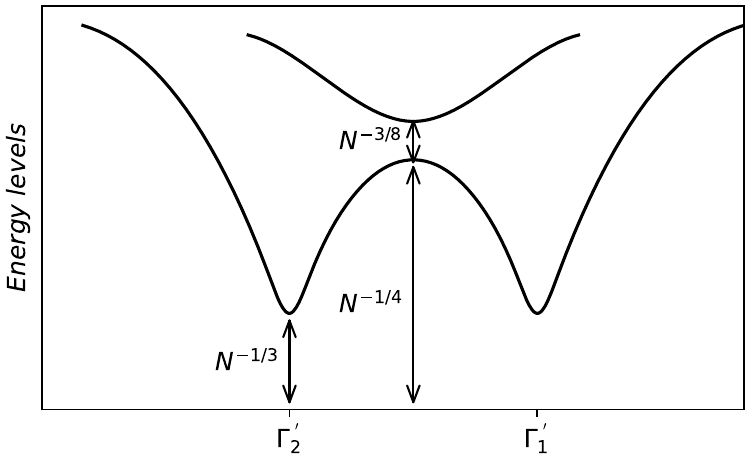}
    \caption{ Lowest two energy levels: without interaction(left), and with interaction (right). The horizontal $\Gamma$ axis is considered to be the ground state energy. \label{fig8} }
\end{figure*}

\section{Conclusions and Outlook} \label{sec_5}
In this work, we have examined the time efficiency of a quantum annealing problem, given the presence of a second-order phase transition along the annealing schedule. The analytical relation between the characteristic annealing time and the scaling exponents was presented in the case of a linear schedule. Specifically, in Sections \ref{sec_3} and \ref{sec_4}, we identified a well-defined second-order phase transition point in the quantum Hopfield model under the dilute-memory condition $p < log_2 N$ by treating the system as a collection of large spins. The relevant scaling exponents ($b = \frac{1}{3}$, $a = \frac{1}{2}$) as well as the annealing time ($T\sim N$) were obtained. 

An interesting direction for future study is the question of how finite connectivity among qubits will affect the system, as it is well-known that large-scale all-to-all connectivity is far from feasible in current quantum annealers. Another important issue is that the Hopfield model itself is not inherently concerned with optimization problems; it is designed to achieve associative memory, so it does not naturally require adiabaticity. This gives great flexibility in choosing a quantum protocol as its dynamical rule. For instance, the quantum algorithms that we mentioned earlier, i.e., quantum approximate optimization algorithms \cite{farhi_2014,farhi_2022}, reverse and iterative cyclic annealing \cite{ohkuwa_2018,wang_2022,zhang_2024}, counter-diabatic driving \cite{berry_2009,sels_2017,claeys_2019,hegade_2022} and quantum walks \cite{callison_2019}—which are designed in the non-adiabatic regime, could serve as fascinating platforms to explore the quantum Hopfield model. These studies may help us better understand how to achieve meaningful quantum associative memory.

\begin{acknowledgments}
This work was supported by the NSF Grants No. DMR-2037654 and No. DMR-2338819.

\end{acknowledgments}

\appendix
\section{Effective quartic potential} \label{ap_a}
Consider first the most probable configuration of $\{ N_l\}$ that all $N_l$ are equal to $N/2^{p-1}$. In this case the $\textbf{T}$ matrix is an identity matrix. As a result, there are exactly $p$ degenerate zero modes (labeled  as $\{b_i\}_{i=1,...,p}$ ) and all other modes are separated by a finite (in the limit $N\to\infty$) gap. Upon orthogonal rotation the quartic Hamiltonian (\ref{s4_H1}) acquires the following form: 
\begin{eqnarray} \label{a_H1_1}
&&\!\!\!\!\!\! H_1 = \frac{1}{4N} \sum_{\mu=1}^{p} \sum_{l,m=1}^{2^{p-1}} K_l^\mu K_m^\mu \left[ \sum_{i=1}^{2^{p-1}} O_{li} (b_i^\dagger +b_i)  \right. \nonumber \\ 
&&\!\!\!\!\!\!\!\!\!\! \left.
\times\! \sum_{j=1}^{2^{p-1}} O_{mj} b_j^\dagger \sum_{k=1}^{2^{p-1}} O_{mk} (b_k^\dagger +b_k) \sum_{q=1}^{2^{p-1}} O_{mq} b_q +  h.c. \right]\!.
\end{eqnarray}
Since the low energy excitations are coming from zero modes, one can drop all massive modes in (\ref{a_H1_1}), i.e., $b_i = b_i^\dagger =0$ for $i > p$. Like for the single pattern case $p=1$, near $\Gamma_c = 1$ one may put $b_i\approx b_i^{\dagger} \approx x_i /\sqrt{2}$ for $ 1\leq i \leq p$, to find for the quartic Hamiltonian in the phase space representation 
\begin{equation} \label{a_H1_2}
H_1 \approx \frac{1}{2N} \sum_{\mu=1}^{p} \sum_{l,m=1}^{2^{p-1}} K_l^{\mu} K_m^{\mu} \sum_{i=1}^{p}O_{li}x_i \left( \sum_{j=1}^{p} O_{mj}x_j \right)^{\!3}.
\end{equation}
Note that the summation limit of $x_i$ has changed to $p$. The remaining column vectors of $\textbf{O}$ in (\ref{a_H1_2}) are eigenvectors of $\textbf{M}$ corresponding to $p$ non-zero eigenvalues. This leads to
\begin{equation}
O^T M O = I \ ,
\end{equation}
with $\textbf{O}$ defined as
\begin{equation} \label{a_O}
O = \frac{K^T}{\sqrt{2^{p-1}}} \ \ \ , \ \ \ (K)_{\mu l} = K^\mu_l \ .
\end{equation}
Applying (\ref{a_O}) to (\ref{a_H1_2}), one finds
\begin{eqnarray} \label{a_H1_3}
&&\!\!\!\!\!\!
 H_1 \approx \frac{1}{2N (2^{p-1})^2}\sum_{\mu,i=1}^{p}\sum_{l,m=1}^{2^{p-1}}(K)_{il}(K^T)_{l\mu}(K)_{\mu m} x_i 
\nonumber \\ 
&&\!\!\!\!\!\!
 \ \ \ \ \ \   \times \left( \sum_{j=1}^{p} (K^T)_{mj}x_j \right)^{\!3}
\nonumber \\ 
&&\!\!\!\!\!\!
= \frac{1}{2N (2^{p-1})^2}\sum_{\mu,i=1}^{p}\sum_{m=1}^{2^{p-1}} 2^{p-1}\delta_{i\mu}(K)_{\mu m} x_i \left( \sum_{j=1}^{p} (K^T)_{mj}x_j \right)^{\!3}
\nonumber \\ 
&&\!\!\!\!\!\!
= \frac{1}{2N 2^{p-1}}\sum_{m=1}^{2^{p-1}}\left[ \sum_{j=1}^{p} (K^T)_{mj} x_j \right]^{\! 4}.
\end{eqnarray}
Since the $(K^{T})_{mj}$'s are either $+1$ or $-1$, only even power terms $x_i^4$ and $x_i^2x_j^2$ contribute for each $m$ in (\ref{a_H1_3}), while odd power terms $x_i x_j^3$, or 
$x_ix_jx_k^2$ cancel each other due to the orthogonality of the $K$ vectors. One thus concludes
\begin{equation} \label{a_H1_4}
    H_1 \approx \frac{1}{2N} \sum_{j=1}^{p} x_j^4 + \frac{3}{2N}\sum_{i\neq j}^{p}x_i^2 x_j^2. 
\end{equation} 
Though (\ref{a_H1_4}) was derived for a specific configuration of $2^{p-1}$ equal size $N_l/2$ spins, it is easy to see that this condition does not matter in the leading order in $1/N$. Indeed, the odd power terms such as $x_ix_j^3$, though non-zero in general, are subleading in $1/N$ due to the random signs of the $K^\mu_l$ elements.

\section{Calculation of $\langle x_k^2 \rangle$ and $\delta \Gamma_k$} \label{ap_b}
In the calculation of expectation values $\langle x_k^2 \rangle$, one encounters the following situations:  
\begin{subequations}\label{b_x2}
    \begin{equation}\label{b_x2_a}
        p^2 + \frac{x^2}{\sqrt{N}} + \frac{x^4}{N} \Rightarrow \omega_a \sim N^{-\frac{1}{4}}, \quad \langle x^2 \rangle \sim N^{\frac{1}{4}};
    \end{equation}
    \begin{equation}\label{b_x2_b}
        p^2 + \frac{x^4}{N} \Rightarrow \omega_b \sim N^{-\frac{1}{3}}, \quad \langle x^2 \rangle \sim N^{\frac{1}{3}};
    \end{equation}
    \begin{equation}\label{b_x2_c}
        p^2 - \frac{x^2}{\sqrt{N}} + \frac{x^4}{N} \Rightarrow \omega_c \sim N^{-\frac{1}{4}}, \quad  \langle x^2 \rangle \sim N^{\frac{1}{2}}, 
    \end{equation}
\end{subequations}
where $\omega$'s represent the energy gap for each case.
For (\ref{b_x2_a}), the quartic part can be ignored for low energy excitations, which are described by the harmonic oscillator. For (\ref{b_x2_c}), the double-well potential can  be regarded as the harmonic oscillator centered around $x_0 \sim\pm N^{1/4} $.

For $\Gamma \approx \Gamma_1$, all modes except mode $k = 1$ are of type (\ref{b_x2_a}), thus the correction to $\Gamma_1$ according to (\ref{s4_deltagamma}) is
\begin{equation}
    \delta \Gamma_1 \sim (p-1)\frac{N^{\frac{1}{4}}}{N} \sim N^{-\frac{3}{4}} \ .
\end{equation}
For $\Gamma \approx \Gamma_2$, mode 1 is of type (\ref{b_x2_c}), modes $k>2$ are of type (\ref{b_x2_a}), and the correction to $\Gamma_2$ is thus
\begin{equation}
    \delta \Gamma_2 \sim \frac{N^{\frac{1}{2}}}{N} + (p-2) \frac{N^{\frac{1}{4}}}{N} \sim N^{-\frac{1}{2}} \ .
\end{equation}
Similarly for $\Gamma \approx \Gamma_k$,
\begin{equation}
    \delta \Gamma_k \sim (k-1)\frac{N^{\frac{1}{2}}}{N} + (p-k)\frac{N^{\frac{1}{4}}}{N} \sim {N^{-\frac{1}{2}}}\ .
\end{equation}
These estimates show that the mean-field renormalized $\Gamma_k$'s are shifted by the amount $\sim {1}/{\sqrt{N}}$. Therefore, the typical spacings between the corrected $\Gamma_k$'s are still of the same order as
those between their bare values. 

\section{Local minimal gaps around critical point}\label{ap_c}
Equation (\ref{H''_4_4}), represents a non-interacting Hamiltonian in the mean-field approximation. A careful reader will notice that such an approximation cannot be applied to those points where energy levels are degenerate. To see this, let us consider the  $p=2$ case. Figure \ref{fig8} shows the basic idea. Without the interaction term, $\sim x_1^2x_2^2/N$, there is a point between $\Gamma_1$ and $\Gamma_2$ that the first excited dynamical state is degenerate. The interaction term opens a gap, $\Delta^{'}$, which can be estimated with the degenerate perturbation theory
\begin{equation}
    \Delta^{'} \sim \frac{\langle 02 | x_1^2  x_2^2|20 \rangle }{N} \sim N^{-{3}/{8}},
\end{equation}
where $| 02 \rangle$ is the state with mode 2  excited, while $| 20 \rangle$ is the state with mode 1 excited, they represent the two energy levels in Fig.~\ref{fig8}. Here the state $|2\rangle$ is the first dynamical state with the even parity. The calculation of this expectation value is a straightforward application of Appendix \ref{ap_b}. Specifically, $\langle 0| x_1^2 | 2\rangle$ is obtained from a double-well Hamiltonian of type (\ref{b_x2_c}). The ground state can be regarded as the superposition of ground states at minima $x_0 \sim \pm N^{\frac{1}{4}}$ such that $|0\rangle = \frac{\psi_0(x_1-x_0)+\psi_0(x_1+x_0)}{\sqrt{2}}$, where $\psi_0$ is the ground state of a single well. The first excited even state can be regarded as the odd superposition of the two locally odd states, i.e., $|2\rangle = \frac{\psi_1(x_1-x_0)-\psi_1(x_1+x_0)}{\sqrt{2}}$, where $\psi_1$ is the first excited state of a single well. Therefore,
\begin{eqnarray} 
&&\!\!\!\!\!\!   \langle 0| x_1^2 | 2\rangle = \frac{1}{2} \int dx_1 x_1^2 \left[ \psi_0(x_1-x_0)+\psi_0(x_1+x_0)\right]   \nonumber \\ 
&&\ \ \ \ \ \ \ \ \ \ \ \ 
\times \left[ \psi_1(x_1-x_0)-\psi_1(x_1+x_0) \right] 
\nonumber \\ 
&& \approx  \frac{1}{2} \int dy \left[ (y+x_0)^2 - (y-x_0)^2 \right] \psi_0(y) \psi_1(y)
\nonumber \\ 
&& =2 x_0 \int dy y \psi_0(y) \psi_1(y)
\sim \frac{x_0}{\sqrt{\omega_c}} \sim N^{{3}/{8}}.
\end{eqnarray}
The matrix element $\langle 0| x_2^2 | 2\rangle$ is easier to calculate, since the related Hamiltonian is of the type (\ref{b_x2_a})
\begin{equation}
    \langle 0| x_2^2 | 2\rangle \sim \frac{1}{\omega_a} \sim N^{{1}/{4}}.
\end{equation}
Finally, it comes to our result $\Delta^{'} \sim N^{-{3}/{8}}$.
As illustrated in the right panel of Fig.~\ref{fig8}, the energy spacing of the ground state and first excited dynamical state, $\sim N^{-1/4}$, is larger than $\Delta^{'}\sim N^{-3/8}$, showing that the avoided crossing transitions at $\Gamma_k^{'}$s are well separated,
and not affected by the interactions terms in Eq.~(\ref{a_H1_4}).

\bibliography{refs}

\end{document}